\documentclass[a4paper,11pt]{article}
\pdfoutput=1
\usepackage{jheppub}
\usepackage{amssymb,amsmath,mathrsfs,enumerate}
\usepackage{graphicx,rotate,multicol}
\usepackage{float}

\usepackage{subfig}

\usepackage{slashed}
\usepackage{mathtools}
\usepackage{multirow}

\allowdisplaybreaks

\title{\boldmath Spinning Primordial Black Holes from First Order Phase Transition}
\author{Indra Kumar Banerjee,}
\author{Ujjal Kumar Dey}
\affiliation{Department of Physical Sciences, Indian Institute of Science Education and Research Berhampur,\\Transit Campus, Government ITI, Berhampur 760010, Odisha, India}

\emailAdd{indrab@iiserbpr.ac.in}
\emailAdd{ujjal@iiserbpr.ac.in}

\abstract{
We conduct a novel study to obtain the initial spin of the primordial black holes created during a first-order phase transition due to delayed false vacuum decay. Remaining within the parameter space consistent with observational bounds, we express the abundance and the initial spin of the primordial black holes as functions of the phase transition parameters. The abundance of the primordial black holes is extremely sensitive to the phase transition parameters. We also find that the initial spin weakly depends on all parameters except the transition temperature.}

\begin{document}
\maketitle
\flushbottom

%%%%%%%%%%%%%%%%%%%%%%%%%%%%%%%%%%%%%%%%%%%%%%%%%%%%%%
%%%%%%%%%%%%%%%%%%%%%%%%%%%%%%%%%%%%%%%%%%%%%%%%%%%%%%
\section{Introduction}
\label{sec:intro}
%%%%%%%%%%%%%%%%%%%%%%%%%%%%%%%%%%%%%%%%%%%%%%%%%%%%%%
%%%%%%%%%%%%%%%%%%%%%%%%%%%%%%%%%%%%%%%%%%%%%%%%%%%%%%
Since its ideation, primordial back holes (PBH)~\cite{Zeldovich:1967lct, Hawking:1971ei, Carr:1974nx, Carr:1975qj} got quite an impressive attention not only as a potential solution to the dark matter problem~\cite{Chapline:1975ojl} but also as a seed of supermassive black holes~\cite{Kawasaki:2012kn} which plays crucial role in galactic dynamics. The discovery of gravitational waves, generated by the black-hole mergers, by the LIGO and Virgo collaborations (LVC)~\cite{LIGOScientific:2016aoc} reinvigorated studies on PBH. 
They are nothing but black holes formed in the early universe by the non-stellar collapse, e.g. collapse from inhomogeneities during radiation~\cite{Carr:1974nx, Green:2000he} and matter dominated era~\cite{Khlopov:1980mg, Nayak:2009pk, Carr:2017edp, Cotner:2017tir}, collapse from inflationary perturbations~\cite{Carr:1993aq,Bullock:1996at,Randall:1995dj,Kim:1999xg,Saito:2008em,Pi:2017gih}, cosmic string loops~\cite{Hawking:1987bn, Matsuda:2005ez, James-Turner:2019ssu} etc.
Recently, different model-(in)dependent first order phase transition (FOPT) driven formation mechanisms have been proposed~\cite{Crawford:1982yz, Kawana:2021tde, Baker:2021nyl, Huang:2022him,Lu:2022paj, Liu:2021svg, Kawana:2022olo, Gouttenoire:2023naa, Lewicki:2023ioy}. 
%
%Some of these are model dependent mechanisms, i.e. the particle content of the model decides whether there will be any creation of PBH during a FOPT~\cite{Kawana:2021tde,Baker:2021nyl,Huang:2022him,Lu:2022paj}, whereas some of these are model independent~\cite{Liu:2021svg,Kawana:2022olo,Gouttenoire:2023naa,Lewicki:2023ioy}. 
%critical collapse~\cite{Niemeyer:1997mt, Kribs:1999bs},
%
%These model independent mechanisms, especially Ref. \cite{Liu:2021svg} was the first to propose the creation of PBH through collapse over overdense regions in the early universe where these regions are formed due to the asynchronous nature of nucleation of true vacuum bubbles during a FOPT. 
%
One interesting scenario utilizes the asynchronous nature of nucleation of true vacuum bubbles during a FOPT that creates overdense regions which subsequently collapses to form PBH~\cite{Liu:2021svg, Kawana:2022olo, Gouttenoire:2023naa, Lewicki:2023ioy}.
In general, the formation of PBH is regulated by a threshold value of average overdensity due to the density perturbation created by curvature perturbation~\cite{Carr:1975qj, Polnarev:2006aa, Musco:2008hv, Harada:2013epa}. These curvature perturbation can be sourced by above-mentioned primordial processes. 
The overdense regions can be characterized by the prescription of compaction function~\cite{Shibata:1999zs, Musco:2018rwt, Escriva:2019nsa, Escriva:2019phb} and peak theory~\cite{Germani:2018jgr, Yoo:2018kvb, Yoo:2020dkz}. In most cases, the PBH mass is shown to be of the order of the mass within the Hubble horizon at the time of creation of the PBH. The present abundance of a PBH population depends on the initial mass of the PBH and the cosmological power spectrum, which can be characterized by its width and amplitude, that eventually leads to peaks, some of which with overdensity high enough for PBH creation. Apart from mass and abundance, the initial spin of the PBH also has huge phenomenological implications as it can cause changes in the Hawking evaporation spectrum~\cite{Dasgupta:2019cae} and superradiant instability~\cite{Calza:2023rjt} which can pave new ways to probe beyond the standard model (BSM) physics~\cite{Arvanitaki:2010sy,Pani:2012vp,Brito:2014wla,Marsh:2015xka,Arvanitaki:2009fg}. 
The natal PBH spin has been studied from the perspective of the peak density of the overdense regions~\cite{DeLuca:2019buf,Harada:2020pzb,Heavens:1988ikr}. Although, in general, the order of magnitude of these initial spin of the PBHs created during the radiation domination is small~\cite{DeLuca:2019buf,Saito:2023fpt}, it can be altered due to various process through out the evolution of the universe~\cite{Jaraba:2021ces,Hofmann:2016yih,Calza:2021czr,Carr:2023tpt,Ferguson:2019slp}. \\
\indent
In this article, we for the first time obtain the initial spin of the PBH created during a FOPT due to the delay in the false vacuum decay. In the purview of peak theory, we use the standard deviation of the overdensity as a function of the FOPT parameters to obtain the initial PBH spin and its distribution as a function of the FOPT parameters.

The article is organized as follows, in Sec.~\ref{sec:delay_decay} we give brief summary of the delayed decay of the false vacuum during a FOPT. In Sec.~\ref{sec:gen_pbhspin} we express the generation of PBH through this process and obtain the abundance and the spin of these PBHs. In Sec.~\ref{sec:pbh_foptparam} we express the dependence of the PBH properties on the FOPT parameters and finally in Sec.~\ref{sec:concl} we summarize and conclude.

%%%%%%%%%%%%%%%%%%%%%%%%%%%%%%%%%%%%%%%%%%%%%%%%%%%%%%
%%%%%%%%%%%%%%%%%%%%%%%%%%%%%%%%%%%%%%%%%%%%%%%%%%%%%%
\section{Delayed Decay of False Vacuum}
\label{sec:delay_decay}
\begin{figure}[t]
\centering
\includegraphics[scale=0.8]{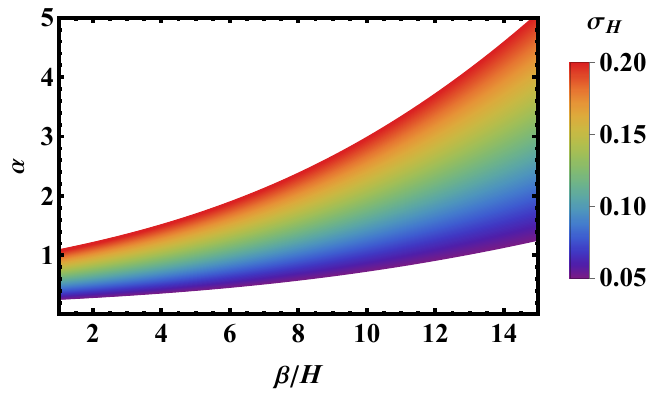}
\caption{$\alpha$ vs. $\beta/H$ for different $\sigma_H$ values.}
\label{alphavsbeta}
\end{figure}

%%%%%%%%%%%%%%%%%%%%%%%%%%%%%%%%%%%%%%%%%%%%%%%%%%%%%%
%%%%%%%%%%%%%%%%%%%%%%%%%%%%%%%%%%%%%%%%%%%%%%%%%%%%%% 
According to a multitude of theories, throughout its evolution our universe has undergone many phase transitions which can have immense implications, especially if they are of first order in nature as they can source gravitational wave backgrounds~\cite{Witten:1984rs, Hogan:1986qda, Gehrman:2023esa, Acuna:2023bkm, Athron:2023xlk, Ellis:2022lft, Banerjee:2023brn}, baryogenesis~\cite{Kuzmin:1985mm, Cohen:1993nk}, primordial black holes~\cite{Crawford:1982yz, Kawana:2021tde, Baker:2021nyl, Huang:2022him, Lu:2022paj, Liu:2021svg, Kawana:2022olo, Gouttenoire:2023naa, Lewicki:2023ioy}, primordial magnetic fields~\cite{Vachaspati:1991nm, Di:2020kbw, Yang:2021uid} etc. Below a certain critical temperature, these first order phase transitions occur through the nucleation of true vacuum bubbles which subsequently expand releasing a huge amount of energy to the bubble walls and the surrounding medium. The nucleation rate of these true vacuum bubbles can be modeled by,
%\begin{equation}
$\Gamma=\Gamma_0 \exp[\beta(t-t_0)]$
%\label{eq:bubnucl}
%\end{equation}
where,
%\begin{equation}
$\beta=-dS(t)/dt\vert_{t=t_0}$
%\label{beta_def}
%\end{equation}
and $S(t)$ is the bounce action of the four or three dimensional instanton solution depending on the temperature of the transition~\cite{Coleman:1977py, Laine:2016hma}.
%\\
%
%\indent
Since the nucleation of the true vacuum bubble is a probabilistic process, it is possible that during a FOPT in some Hubble region, the bubble nucleation is delayed, which can lead to a region with overdensity $\delta$($=\delta\rho/\rho$) that can collapse to form a PBH~\cite{Liu:2021svg,Kawana:2022olo,Gouttenoire:2023naa,Lewicki:2023ioy}. In order to obtain this overdensity in terms of the FOPT strength ($\alpha$), inverse duration ($\beta$), and temperature ($T$), we use the relevant equations of motions and the Friedmann's equation\footnote{Throughout this article, we have considered $8\pi G=c^2=1$.},
\begin{gather}
H^2=\dfrac{\rho_V + \rho_r + \rho_w}{3},
\label{eq:friedmann}
\\
\rho_V=F(t)\Delta V,
\label{eq:vac_dens}
\\
\dfrac{d(\rho_r + \rho_w)}{dt}+4H(\rho_r + \rho_w)=-\dfrac{d\rho_V}{dt},
\label{eq:energy_exchange}
\end{gather}
where $F(t)$ is the average probability of the false vacuum~\cite{Turner:1992tz}, $\Delta V=\alpha\rho_{r0}$ with $\rho_{r0}$ being the radiation energy density at the nucleation temperature; $\rho_V$, $\rho_r$ and $\rho_{w}$ are the vacuum, background radiation and the bubble wall energy density. 
%\\
%\indent
For the peak theory study, the standard deviation of the overdensity ($\sigma_H$) in a Hubble-sized region can be obtained by solving these equations. Using the analysis of~\cite{Liu:2022lvz}, we show the dependence of the standard deviation of overdensity on $\alpha$ and $\beta/H$ in Fig.~\ref{alphavsbeta} which is in the similar spirit of the Fig. 2 of Ref.~\cite{Liu:2022lvz}.\footnote{Note that Ref.~\cite{Liu:2022lvz} denotes the standard deviation of overdensity as $\delta_H$ whereas we denote it as $\sigma_H$ in order to be consistent with the notation in the Press-Schechter formalism and peak theory.}
The dependence of $\sigma_H$ on $\alpha$ and $\beta/H$ can be seen from the above figure, i.e. $\sigma_H$ increases with $\alpha$ and decreases with $\beta/H$. This behaviour shows that the curvature perturbation is higher for high $\alpha$ and low $\beta/H$ values. This is consistent with the existing understanding, i.e., the higher $\alpha$ (stronger FOPT) and lower $\beta/H$ (slower FOPT) leads to better production of PBHs. This is because higher $\alpha$ value leads to higher release in energy whereas lower $\beta/H$ signifies a slow FOPT which in turn allows the overdense region to reach the threshold for collapse.
The overdense regions created from the delayed decay of the false vacuum produces a curvature perturbation. The power spectrum of this curvature perturbation and $\sigma_H$ during the delay of vacuum decay is related as~\cite{Liu:2022lvz},
\begin{align}
\mathcal{P_R}(k)=34.5[\sigma_H(\alpha,\beta/H)]^2(k/ \mathcal{H})^3,
\label{eq:curv_powerspect}
\end{align}
where $\mathcal{H} (=aH)$ is the comoving Hubble parameter. It also worth mentioning that since only the delay of false vacuum decay in a Hubble-sized region is considered, $k\geq \mathcal{H}$ is neglected as they correspond to scales smaller than that of a Hubble region.

%%%%%%%%%%%%%%%%%%%%%%%%%%%%%%%%%%%%%%%%%%%%%%%%%%%%%%
%%%%%%%%%%%%%%%%%%%%%%%%%%%%%%%%%%%%%%%%%%%%%%%%%%%%%%
\section{Generation of Spinning PBH}
\label{sec:gen_pbhspin}
%%%%%%%%%%%%%%%%%%%%%%%%%%%%%%%%%%%%%%%%%%%%%%%%%%%%%%
%%%%%%%%%%%%%%%%%%%%%%%%%%%%%%%%%%%%%%%%%%%%%%%%%%%%%%
Various processes in the early universe create curvature perturbation (or density perturbation which then create curvature perturbation). Depending on the amplitude of the perturbation, they either dissipate in the background or collapse to form  PBHs~\cite{Niemeyer:1999ak,Musco:2004ak,Escriva:2021aeh}. As mentioned in the previous section, delayed vacuum decay during a FOPT can lead to overdense regions and the curvature perturbation associated to that is given in Eq.~\eqref{eq:curv_powerspect}.\\
\indent
Using the peak theory formalism, we express the zeroth spectral moment of the curvature perturbation as~\cite{DeLuca:2019buf,Escriva:2022duf,Yoo:2018kvb,Young:2019yug},
\begin{align}
\sigma^2&=\dfrac{16}{81}\int \dfrac{dk}{k}\mathcal{P_R}(k)\left(\dfrac{k}{\mathcal{H}}\right)^{4}W^2(k)T^2(k),
\label{eq:momnents}
\end{align}
where $W(k)$ and $T(k)$ are the window functions and the transfer functions respectively. In this article we use a top-hat window function instead of a gaussian smoothing function as the later changes the shape of the perturbation significantly. However, the top-hat function is not efficient enough to remove the small scale perturbations and therefore a transfer function has been used as well. Typically these functions can be expressed as,
\begin{align}
W(k)&=3\dfrac{\sin(kr_m)-kr_m\cos(kr_m)}{(kr_m)^3}, \\
T(k)&=3\dfrac{\sin(kr_m/\sqrt{3})-(kr_m/\sqrt{3})\cos(kr_m/\sqrt{3})}{(kr_m/\sqrt{3})^3},
\end{align}
where $r_m$ is the smoothing scale which has been taken to be the horizon scale. This is due to the fact that contributions from $k> \mathcal{H}$ do not contribute to the formation of PBH~\cite{Liu:2022lvz}\\
\indent
It is to be noted that though the curvature perturbation has been assumed to be gaussian, the overdensity is not necessarily gaussian in nature. Therefore, following the prescription in Ref.~\cite{Young:2019yug}, we consider the gaussian component of the overdensity $\delta_l$ which is related to the overdensity as $\delta=\delta_l-3\delta_l^2/8$, where $\delta$ is the total overdensity in the CMC gauge. Furthermore, the threshold of the gaussian component of the overdensity can be connected with the threshold of the total overdensity as~\cite{Young:2019yug},
\begin{align}
\delta_{c,l \pm}&=\dfrac{4}{3}\left(1 \pm \sqrt{\dfrac{2-3\delta_c}{2}} \right),
\end{align}
where $\delta_c$ is the threshold for the total oversdensity. It is to be noted that we consider $\delta_{c,l -}$ as the threshold for $\delta_l$ as the other solution corresponds to a type-II perturbation and its value remains beyond the threshold. Now we move on to the calculation of the abundance and the spin of the PBH population originating from this setup.
\subsection{PBH Abundance}
\begin{figure}[t]
\centering 
\includegraphics[scale=0.4]{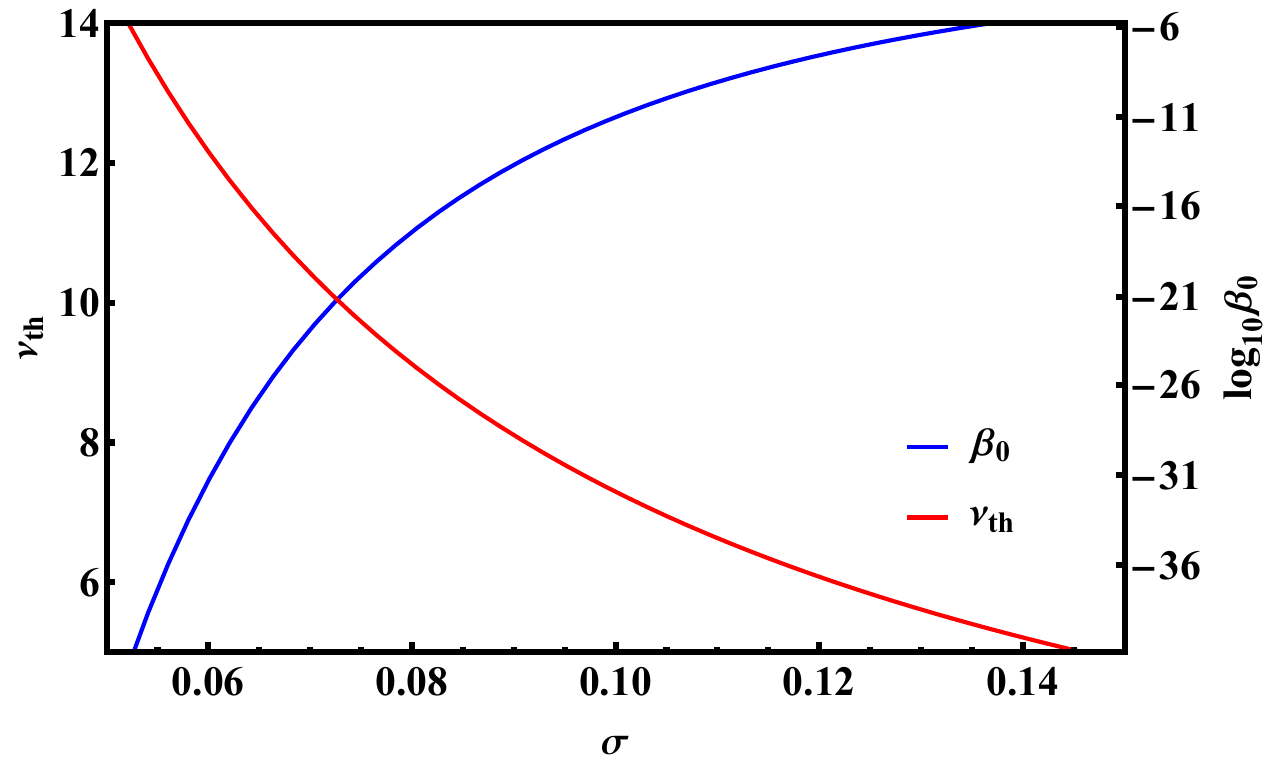}
\caption{\footnotesize Behaviour of $\nu_{\mathrm{th}}$ and $\beta_0$ with respect to $\sigma$. In both the cases $\delta_{c}$ has been taken as 0.535.}
\label{sigmadependence}
\end{figure}

\begin{figure}[t]
\centering 
\includegraphics[scale=0.8]{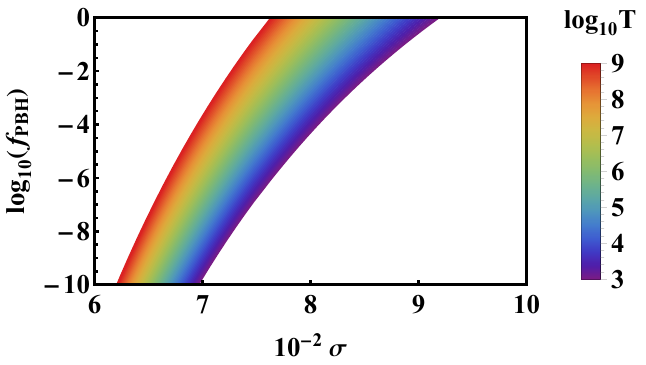}
\caption{\footnotesize $f_{\mathrm{PBH}}$ dependence on $\sigma$ for FOPT temperature in the range $T\in (10^3, 10^9)$ GeV. In all the cases, $\delta_{c}$ has been taken as 0.535.}
\label{sigmadependencef}
\end{figure}

The PBH mass can be expressed in terms of the threshold of the overdensity, the horizon mass as~\cite{Young:2019yug},
\begin{align}
M_{\mathrm{PBH}}=\mathcal{K}M_H(\delta-\delta_c)^{\gamma^{\prime}}=\mathcal{K}M_H(\delta_l-3\delta_l^2/8-\delta_c)^{\gamma^{\prime}},
\label{mpbh}
\end{align}
where $\mathcal{K}$ is a numerical prefactor which depends on the threshold of the overdensity, and $\gamma^{\prime}$ is taken to be $\sim 0.36$.
To estimate the PBH abundance we note that the fraction of the universe which collapse into PBHs can be calculated as a function of $\nu$ or $\sigma$ as~\cite{Young:2019yug},
%\begin{widetext}
\begin{align}
\beta_0 = \int^{4/3\sigma}_{\nu_\mathrm{th}} d\nu \dfrac{\mathcal{K}}{3\pi}\left(\nu\sigma-\dfrac{3}{8}(\nu\sigma)^2-\delta_c\right)^{\gamma^{\prime}}\left(\dfrac{\mu}{\mathcal{H}\sigma}\right)^3\exp\left(-\dfrac{\nu^2}{2}\right),
\end{align}
%\end{widetext}
where $\nu=\delta_l/\sigma$, $\nu_{\mathrm{th}}=\delta_{c,l-}/\sigma$, and $\mu$ can be expressed as,
\begin{align}
\mu^2=\dfrac{16}{81}\int^{\infty}_0 \dfrac{dk}{k}(k r_m)^4 W^2(k,r) T^2(k,r_m)\mathcal{P}_{\mathcal{R}}(k)k^2.
\end{align}
It is to be noted here that the threshold value $\delta_c$ depends on the shape of the power spectrum as mentioned in Ref.~\cite{Musco:2020jjb}. For the power spectrum we have considered, the value of the threshold $\delta_c=0.535$. It is also worth mentioning that for this threshold value, one can obtain $\mathcal{K}=5.5$ following the prescription of Ref.~\cite{Musco:2020jjb}.
We show the dependence of $\nu_{\mathrm{th}}$ and $\beta_0$ on $\sigma$ in Fig.~\ref{sigmadependence} which shows that $\nu_{\mathrm{th}}$ decreases with $\sigma$ and as a consequence $\beta_0$ increases exponentially.
%\\
%%
%\indent
%
Therefore, if $\sigma$ is small $\nu_{\mathrm{th}}$ is very high, which in turn makes $\beta_0$ extremely small, i.e. a very small patch of the entire universe reaches the threshold value of 0.535 if the standard deviation of the overdensity distribution is very small. This is consistent with the assumption that overdensity is a statistical variable with a gaussian component. This gaussian aspect also suggests that if the standard deviation is low then probability that some part of the universe reaches the threshold overdensity is very low. To illustrate this further, we can see that for $\sigma=0.06$, $\nu_{\mathrm{th}}=13$ and $\beta_0\sim 10^{-31}$. So, as $\sigma$ goes further below, still some fraction of the universe can achieve the threshold overdensity, but it will be too low to eventually create a significant abundance of primordial black holes.
The total energy fraction of the universe contained in the PBH can be expressed as,
\begin{align}
\Omega_{\mathrm{PBH}}=\int^{M_{\mathrm{max}}}_{M_{\mathrm{min}}} \mathrm{d}(\mathrm{ln}M_{H})\left(\dfrac{M_{\mathrm{eq}}}{M_{H}}\right)^{1/2}\beta_0,
\end{align}
where $M_{\mathrm{max}}$ and $M_{\mathrm{min}}$ are the maximum and the minimum horizon masses at which the PBHs form which can be found out from Eq.~\eqref{mpbh}, and $M_{\mathrm{eq}}$ is the horizon mass at the matter-radiation equality. The final abundance can be calculated as,
\begin{align}
f(M_{\mathrm{PBH}})=\dfrac{1}{\Omega_{\mathrm{CDM}}}\dfrac{\mathrm{d}\Omega_{\mathrm{PBH}}}{\mathrm{d}(\mathrm{ln}M_{\mathrm{PBH}})}.
\end{align}
For simplicity, we consider the PBHs which form with a mass approximately equal to the horizon mass, i.e.
\begin{align}
M_{\mathrm{PBH}}=9.23\times 10^{31}\left(\dfrac{T}{\mathrm{GeV}}\right)^{-2}\mathrm{~g}
\label{mpbhT}
\end{align}
where $T$ is the transition temperature and we have taken the density of dark mater $\Omega_{\mathrm{CDM}}=0.26$~\cite{Planck:2018vyg} and the effective degrees of freedom $g_{*}\sim 100$ which is valid for our parameter ranges. To understand the behavior of $f_{\mathrm{PBH}}$, we show its dependence on $\sigma$ for different values of $T$ in Fig.~\ref{sigmadependencef}. It is evident that, like $\beta_0$, $f_{\mathrm{PBH}}$ is also extremely sensitive on $\sigma$. It is also to be noted that as $T$ increases, the $f_{\mathrm{PBH}}$ increases for the same value of $\sigma$.
\subsection{PBH Spin}
\begin{figure}[t]
\centering 
\includegraphics[scale=0.4]{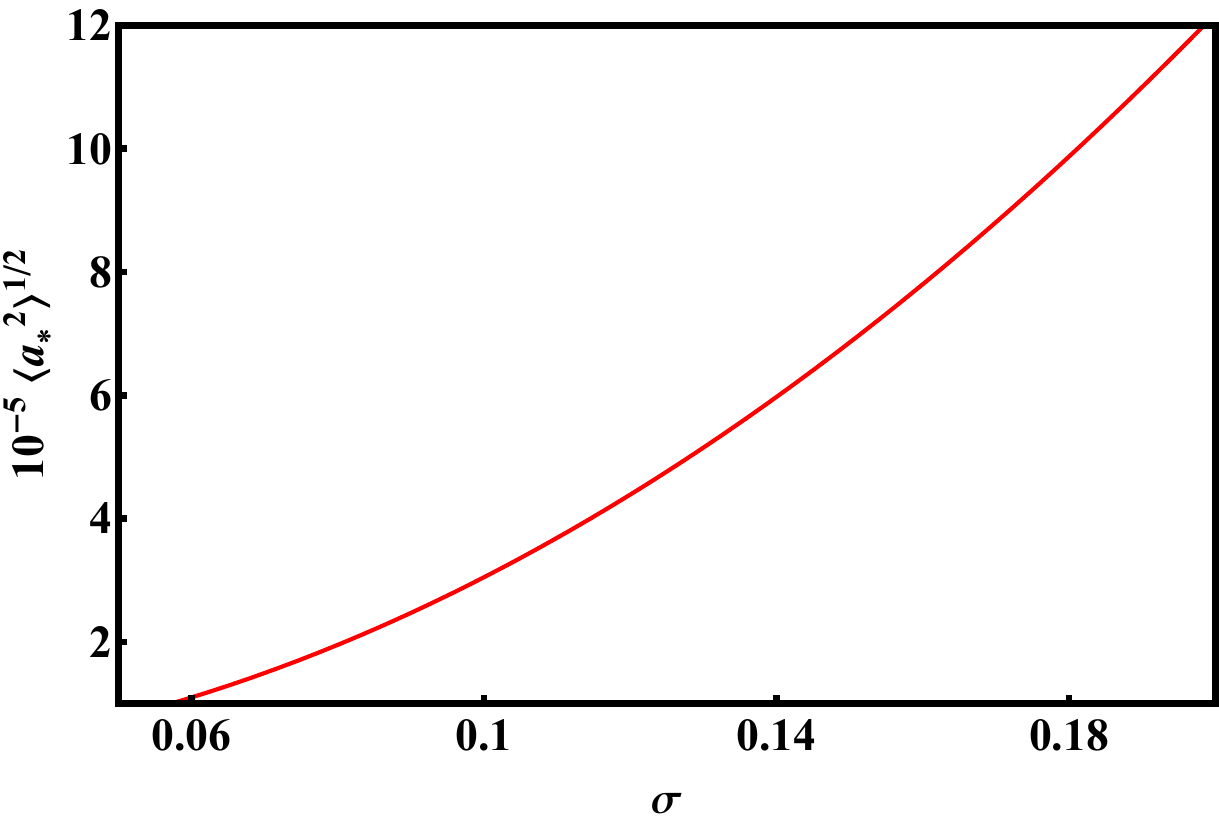}
\caption{\footnotesize $\langle a_*^2\rangle^{1/2}$ dependence on $\sigma$. In all the cases $\delta_c$ has been taken as 0.535.}
\label{sigmadependencea}
\end{figure}

Now we come to the pivotal point of this article i.e., the spin of the PBHs. First, we briefly mention how spin can be generated from the perspective of peak theory assuming the overdensity profile to have a gaussian component. For non-monochromatic power spectrum of the curvature perturbation, the peak of a gaussian overdensity component can have a slightly non-spherical behavior depending on the amount of deviation from monochromaticity. Therefore, ellipsoidal perturbations are considered to characterize the overdense peaks which eventually collapse to form PBHs. If the inertia tensor of the ellipsoidal perturbation is not aligned with the velocity shear, then the collapsing region can have a non-zero spin. In the radiation dominated universe, which is the most-likely period of the FOPTs, this spin generation is meager, as the deviation from the spherical symmetry is very less in radiation domination. Still it is important to characterize this spin as there exists several mechanisms responsible for the spin-up of a PBH~\cite{Jaraba:2021ces,Hofmann:2016yih,Calza:2021czr}.
%\\
%
Considering the first order anisotropy around the overdensity peak, the distribution of the spin parameter can be estimated from peak theory. Following Eq.~(6.30) of Ref.~\cite{DeLuca:2019buf} we express the distribution of the normalized spin parameter $h$,
\begin{align}
P(h)dh =\exp\big[-2.37-4.12\left(\ln h\right) -1.53\left(\ln h\right)^2-0.13\left(\ln h\right)^3\big]dh,
\label{eq:adistvsa}
\end{align}
where $h$ is related to the dimensionless Kerr parameter $a_{*}$ as~\cite{Escriva:2022duf}\footnote{In our work, we consider a broad power spectrum and hence the relation between $h$ and $a_*$ is different from the one in Ref.~\cite{Escriva:2022duf}.},
\begin{align}
h=5.62\times 10^2\dfrac{ a_{*}\nu^2}{\sqrt{1-\gamma^2}}.
\label{eq:ahconv}
\end{align}
In this we consider the PBH masses very close to the horizon mass. 
Here the parameter $\gamma$, whose deviation from unity signifies non-monochromaticity, can be expressed in terms of the spectral moments,
%\begin{align}
$\gamma=\sigma_1^2/ (\sigma \sigma_2)$, where $\sigma_1$ and $\sigma_2$ are the first and the second spectral moment respectively.
%\end{align}
Typically the value of $\gamma$ lies in the range $(0.85,1)$~\cite{DeLuca:2019buf,Escriva:2022duf,Harada:2020pzb}. In our case, from  Eq.~\eqref{eq:momnents} we get $\gamma \approx 0.96$.
In Ref.~\cite{Harada:2020pzb} the value of the initial spin was calculated for a narrow power spectrum, however, that is not the case here. Therefore, we calculate the initial spin for a broad power spectrum, i.e. the one mentioned in Eq.~\eqref{eq:curv_powerspect}. In doing so we take into account all scales on or above the horizon. However, we do not consider scales smaller than the horizon scale as they do not contribute to the creation of PBHs. Thus we obtain the variance of the dimensionless Kerr parameter as,
\begin{align}
\label{eq:redKerrPar}
\langle a_*^2\rangle^{1/2} = 1.09\times 10^{-2} \left(\dfrac{M_{\mathrm{PBH}}}{M_H}\right)^{-1/3}\left(\dfrac{1}{\nu}\right)^{-2}.
\end{align}
We reiterate that in the above expression the numerical prefactor has been derived for a broad power spectrum, whereas, in Ref.~\cite{Harada:2020pzb}, the analogous numerical prefactor can obtained as $0.115$, which is more than ten times higher than the one we obtain. Hence, we conclude that broad power spectrum lead to lower initial spin of PBH.
It is evident that $T$-dependence on $\langle a_*^2\rangle^{1/2}$ identically cancels out as we consider $M_{\mathrm{PBH}}\sim M_H$. In Fig.~\ref{sigmadependencea}, we have shown the dependence of $\langle a_*^2\rangle^{1/2}$ on $\sigma$. It is to be noted here that the rms value of the initial spin of the PBHs does not depend on the FOPT temperature as $\sigma$ is only a function of $\alpha$ and $\beta/H$ and not $T$. It can be seen from Fig.~\ref{sigmadependencea} that $\langle a_{*}^2\rangle^{1/2}$ increases with $\sigma_H$. However, to be in the physically realistic domain, we only consider values of $\sigma$ which will not violate any abundance constraints on PBH. To illustrate this further, consider e.g., for $\sigma\approx 0.062$ and $T=10^9\mathrm{~GeV}$, $f_{\mathrm{PBH}}\sim 10^{-10}$ and $\langle a_{*}^2\rangle^{1/2}\sim 10^{-5}$, but for the same value of FOPT temperature, at $\sigma\approx 0.072$, $f_{\mathrm{PBH}}\sim 1$ and $\langle a_{*}^2\rangle^{1/2}\sim 1.6\times 10^{-5}$.

Therefore, though for this FOPT temperature, higher values of $\sigma$ will lead to higher values of spin, but that will also lead to overproduction of PBHs. Hence, throughout our study, we confine ourselves to the physically realistic parameter space and investigate the spin of PBHs, and we take this into account while choosing our benchmark parameters.

%%%%%%%%%%%%%%%%%%%%%%%%%%%%%%%%%%%%%%%%%%%%%%%%%%%%%%
%%%%%%%%%%%%%%%%%%%%%%%%%%%%%%%%%%%%%%%%%%%%%%%%%%%%%%
\section{Dependence on FOPT Parameters}
\label{sec:pbh_foptparam}
%%%%%%%%%%%%%%%%%%%%%%%%%%%%%%%%%%%%%%%%%%%%%%%%%%%%%%
%%%%%%%%%%%%%%%%%%%%%%%%%%%%%%%%%%%%%%%%%%%%%%%%%%%%%%
We now present the main results, i.e. the effect of the the FOPT parameters, e.g., $\alpha$, $\beta/H$, and $T$ on the abundance and spin of the PBHs.
\begin{figure}[H]
\begin{center}
\subfloat[\label{alphadepf}]{\includegraphics[scale=0.27]{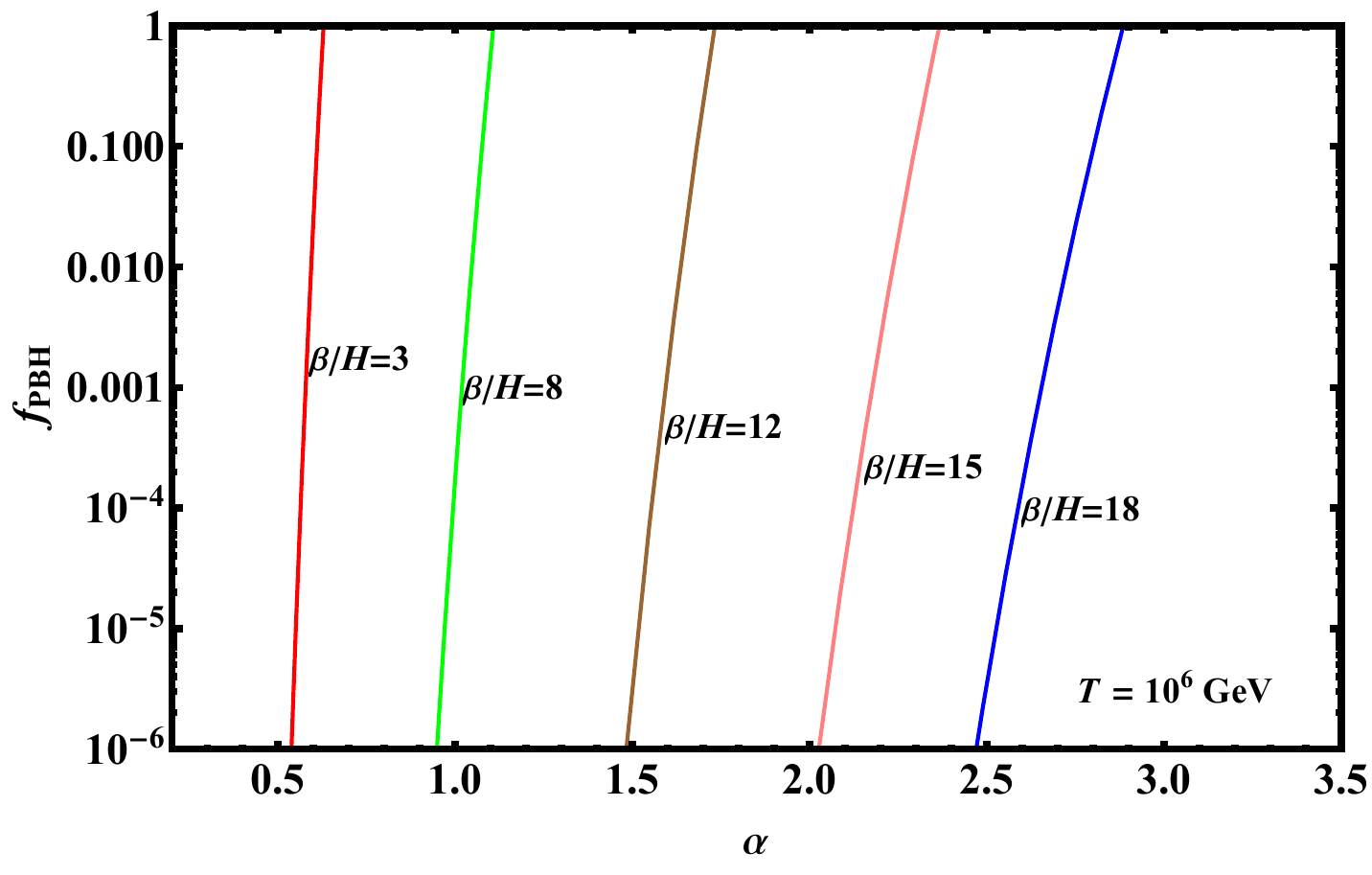}} \qquad \qquad
\subfloat[\label{alphavsa}]{\includegraphics[scale=0.27]{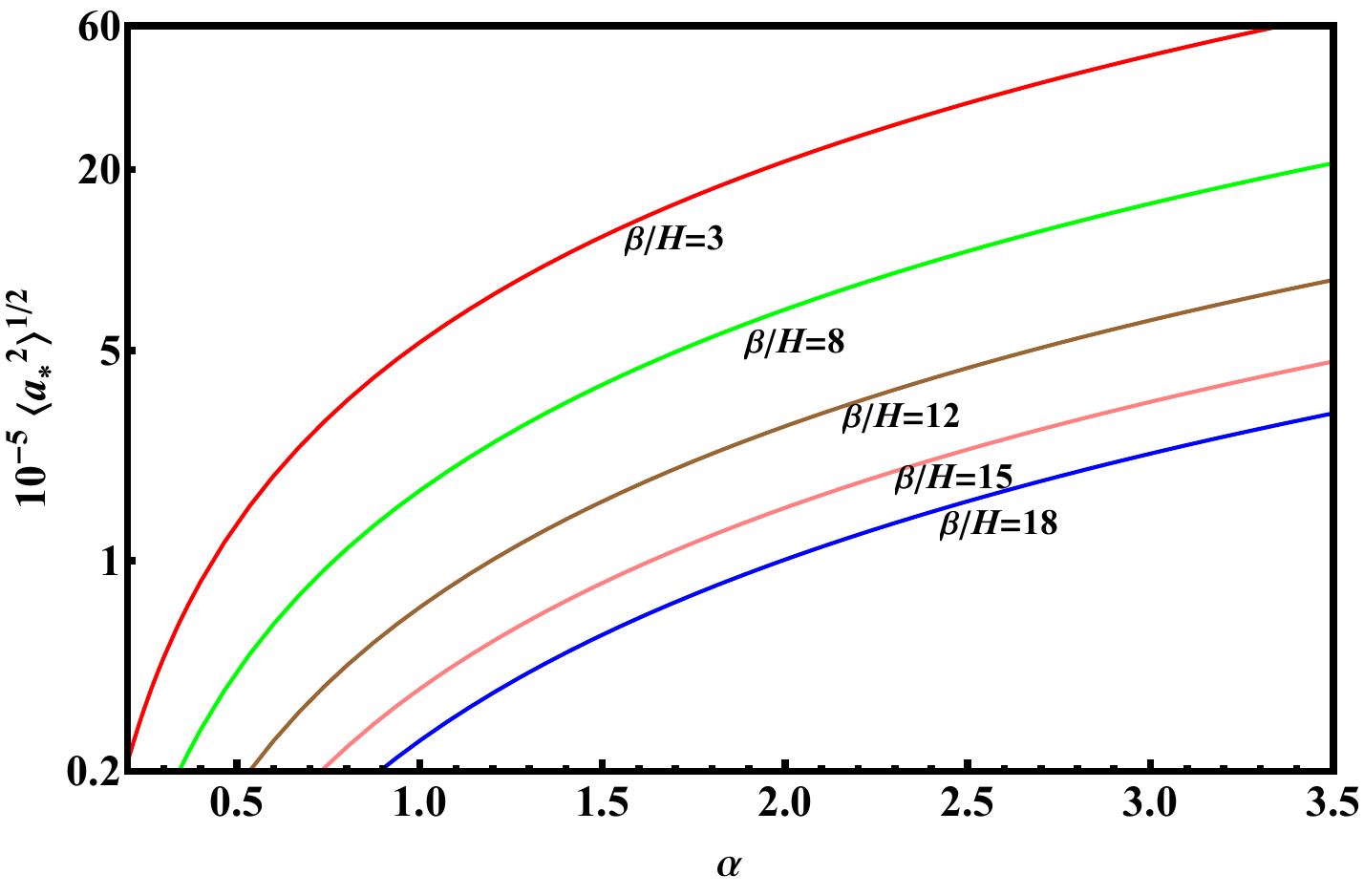}}
\\%
\subfloat[\label{betadepf}]{\includegraphics[scale=0.27]{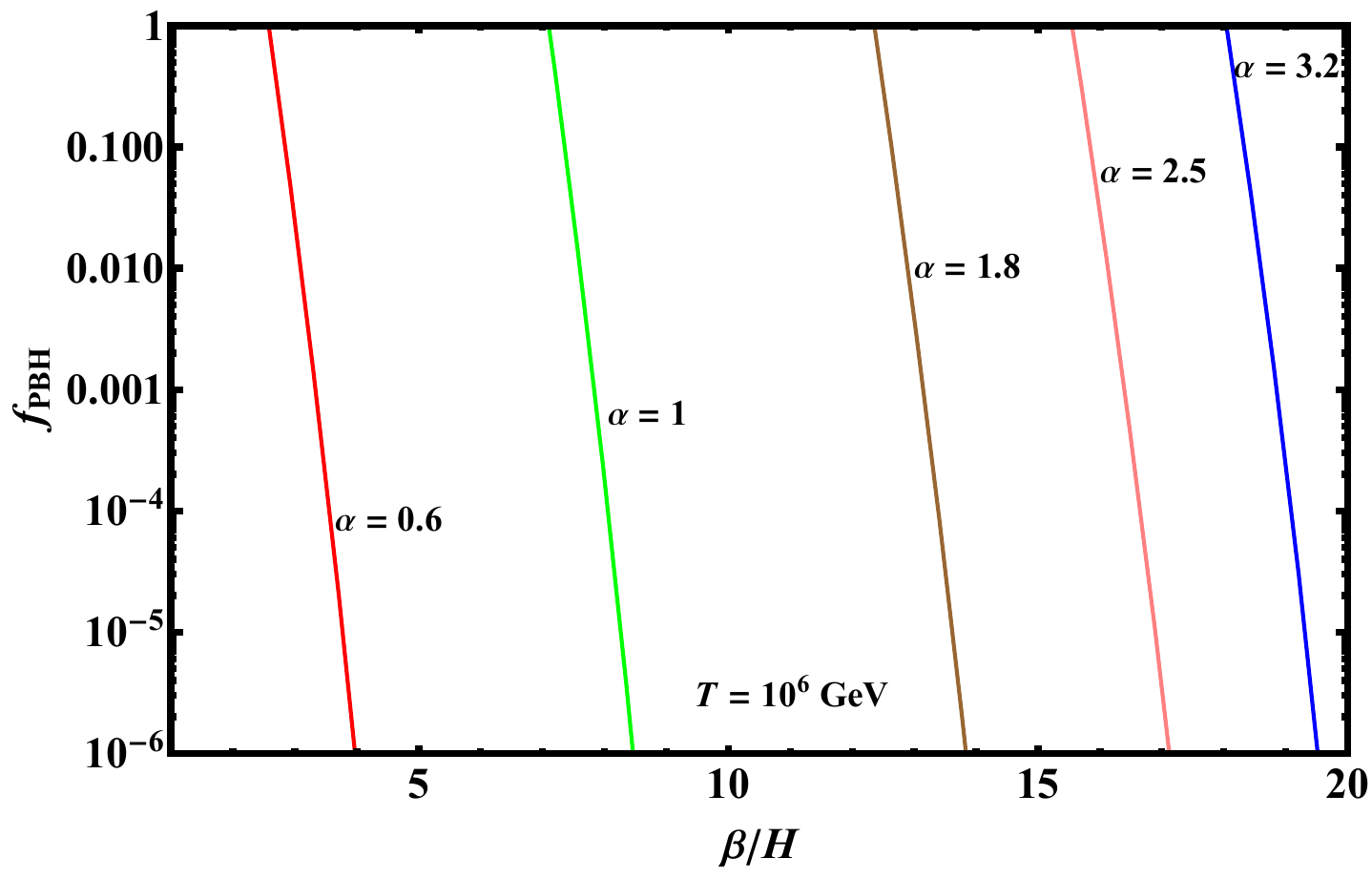}} \qquad \qquad 
\subfloat[\label{betavsa}]
{\includegraphics[scale=0.265]{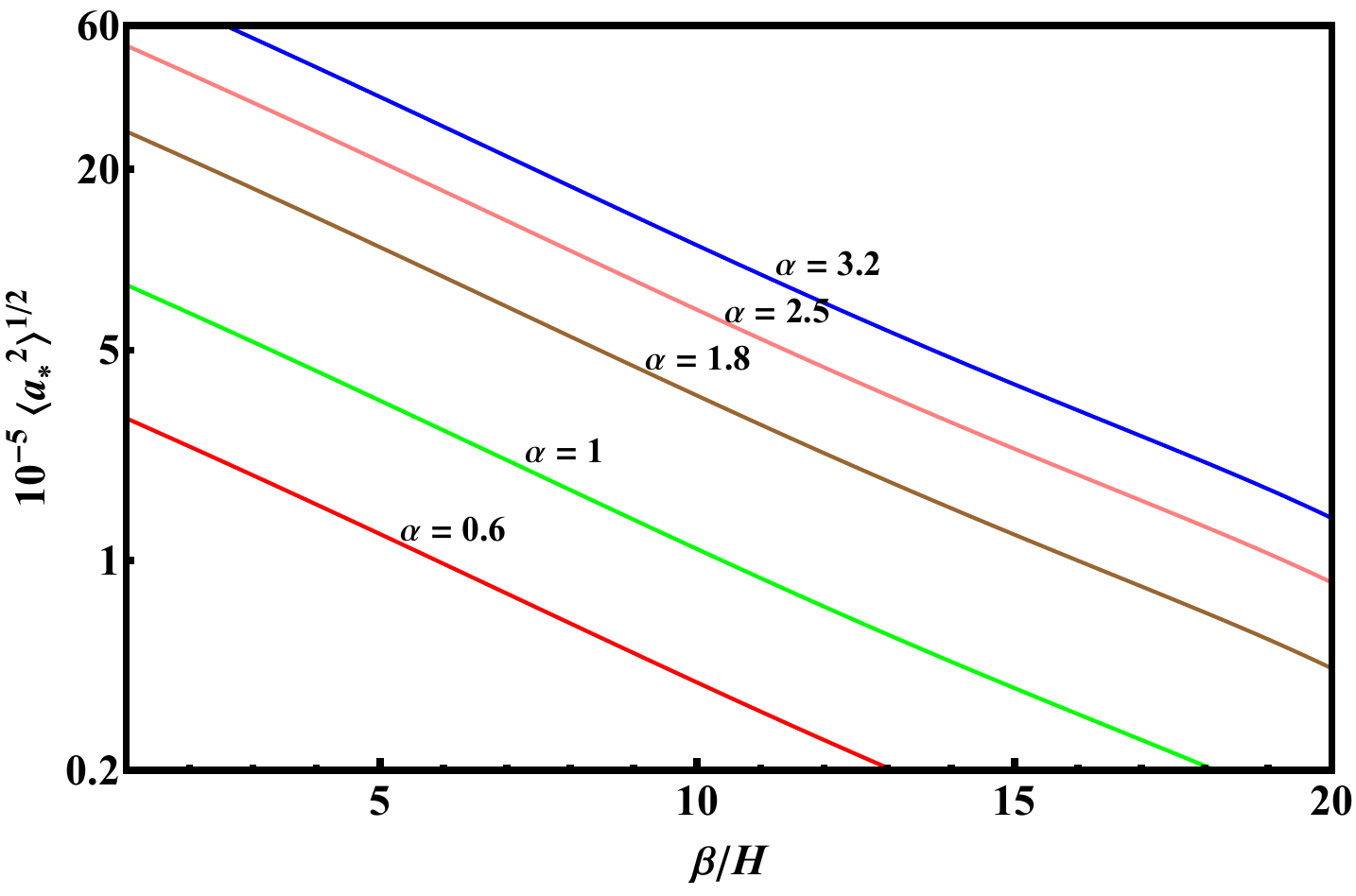}}
\caption{\footnotesize Dependence of (a) $f_{\mathrm{PBH}}$, and (b) $\langle a_*^2\rangle^{1/2}$ on $\alpha$ for various $\beta/H$.  Dependence of (c) $f_{\mathrm{PBH}}$, and (d) $\langle a_*^2\rangle^{1/2}$ on $\beta/H$ for various $\alpha$. We take an illustrative $T = 10^6$ GeV to present the $f_{\mathrm{PBH}}$ cases.
}
\label{Fig3}
\end{center}
\end{figure}
Starting with $\alpha$, in Figs.~\ref{alphadepf} and \ref{alphavsa} we show the dependence of $f_{\mathrm{PBH}}$ and $\langle a_{*}^2\rangle^{1/2}$ on $\alpha$ for different $\beta/H$, respectively. We have used a FOPT temperature of $10^6\mathrm{~GeV}$. Similar qualitative features are expected at other temperatures too. It is also evident from these plots that $f_{\mathrm{PBH}}$ changes very steeply with $\alpha$ as opposed to $\langle a_{*}^2\rangle^{1/2}$, though both of them increases with $\alpha$ for a fixed $\beta/H$ value. It can also be noted that for specific $\alpha$, $\beta/H$ combinations, $f_{\mathrm{PBH}}\sim 1$ and $\langle a_{*}^2\rangle^{1/2}\sim 2\times 10^{-5}$. This shows that for these kind of processes, the generated initial spin of the PBH for most of the PBH population is quite low, which in a way is consistent with the PBH produced in the radiation dominated era~\cite{DeLuca:2019buf}.
The behaviors of $f_{\mathrm{PBH}}$ and $\langle a_{*}^2\rangle^{1/2}$ with varying $\beta/H$ for different values of $\alpha$ are shown in Figs.~\ref{betadepf} and \ref{betavsa} respectively, where it can be seen that though both $f_{\mathrm{PBH}}$ and $\langle a_{*}^2\rangle^{1/2}$ decreases with increasing $\beta/H$, the change is much steeper in case of $f_{\mathrm{PBH}}$ in comparison with $\langle a_{*}^2\rangle^{1/2}$. 
Since $\langle a_{*}^2\rangle^{1/2}$ is independent of $T$, we show the $T$-dependence of $f_{\mathrm{PBH}}$ in the left panel of Fig. \ref{fvsTdep}. It can be understood from the figure that different combinations of $\alpha$ and $\beta/H$ leads to different abundances of PBHs depending on the FOPT temperatures. To illustrate this further, for $(\alpha,\beta/H)=(2.3,15)$, a FOPT at $10^5\mathrm{~GeV}$ leads to PBH of abundance $f_{\mathrm{PBH}} \sim 0.008$, which narrowly escapes the Subaru HSC bound~\cite{Niikura:2017zjd}, whereas for the same combination, a FOPT at $9\times 10^6\mathrm{~GeV}$ gives $f_{\mathrm{PBH}} \sim 1$. This gives an overall idea about the three essential FOPT parameters, i.e. $\alpha$, $\beta/H$, and $T$ for realistic PBH abundance.\\
\indent
The PBH spin distribution can be depicted by the right panel of Fig.~\ref{fvsTdep} where we find that it just gets slight shift depending on the FOPT parameters. 
We reiterate that since the spin of a PBH does not depend on the FOPT temperature, the only relevant parameters are $\alpha$ and $\beta/H$ and moreover we use only those values that gives rise to allowed PBHs. \\
\indent
\begin{figure}[t]
\begin{center} 
\includegraphics[scale=0.32]{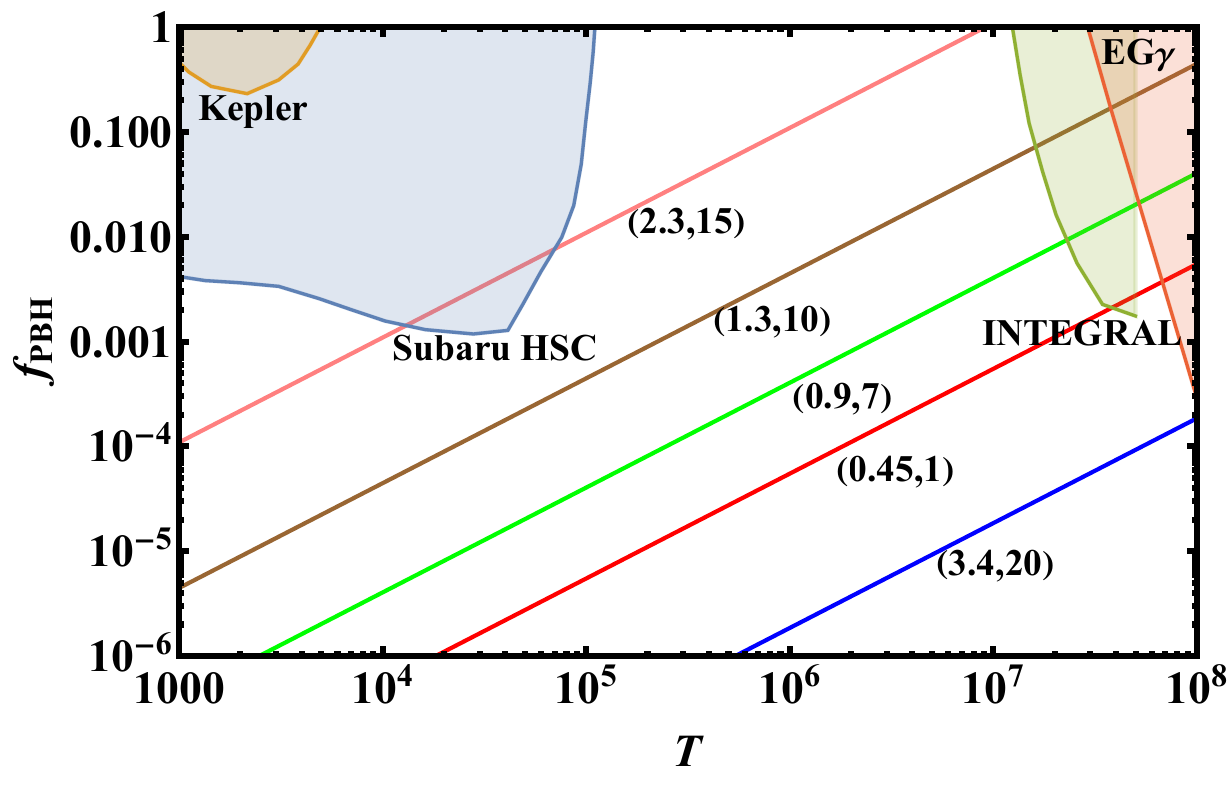}\quad \quad \quad
\includegraphics[scale=0.3]{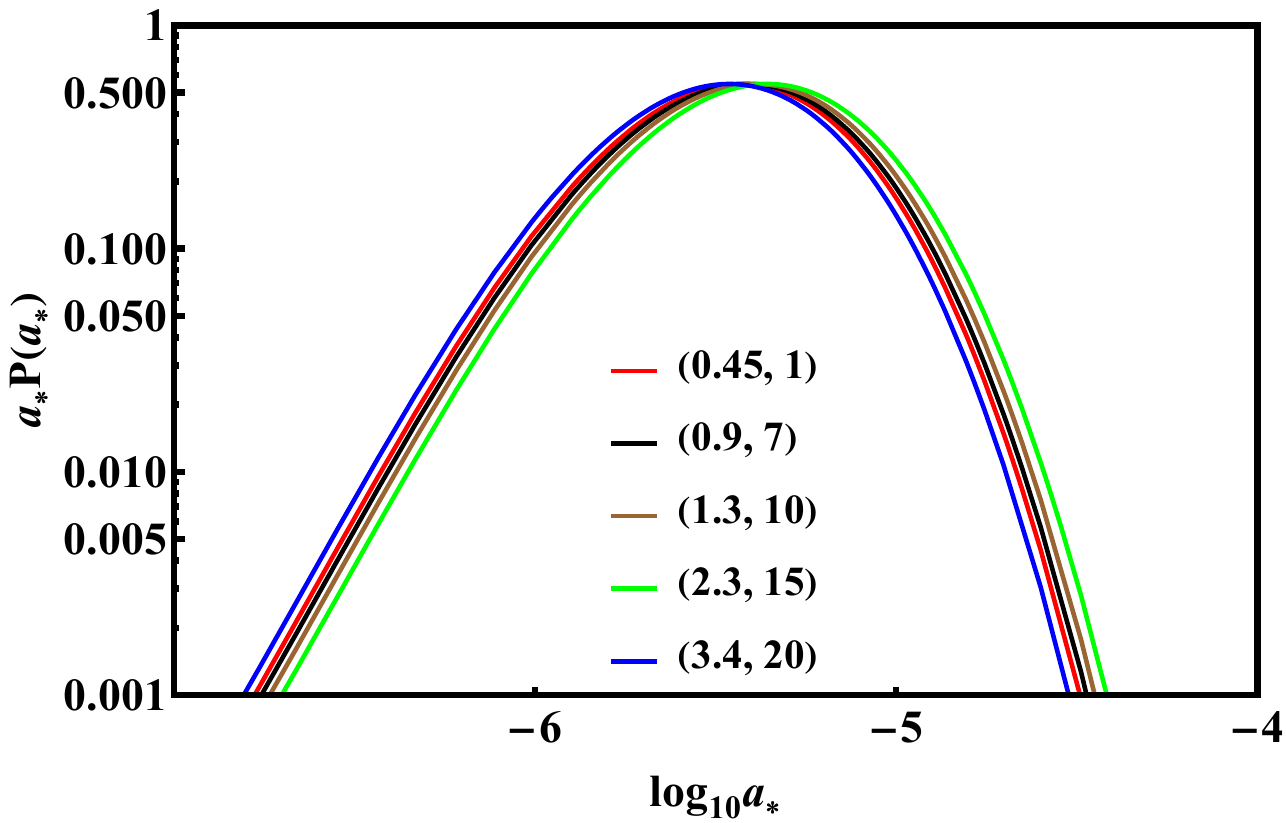}
% 2.5\times10^{-4}
\caption{\footnotesize (left) Dependence of $f_{\mathrm{PBH}}$ on $T$ for different combinations of ($\alpha$, $\beta/H$). The abundance constraints on $f_{\mathrm{PBH}}$ from Kepler\cite{Griest:2013esa}, Subaru HSC~\cite{Niikura:2017zjd}, EG$\gamma$~\cite{Carr:2009jm} and INTEGRAL~\cite{DeRocco:2019fjq,Laha:2019ssq} has been shown. It is to be noted that these constraints are usually expressed in terms of PBH mass and we have converted them to the FOPT temperature domain using Eq.~\eqref{mpbhT}. (right) The distribution of $a_{*}$ for different combinations of ($\alpha$, $\beta/H$).
}
\label{fvsTdep}
\end{center}
\end{figure}
For completeness, now we consider a few specific benchmark cases of FOPTs which are motivated by BSM scenarios, e.g., the Peccei-Quinn axion model (I)~\cite{Peccei:1977ur}, Randall-Sundrum dilaton-like scenario (II)~\cite{Randall:2006py}, SM effective field theory (III)~\cite{Huber:2007vva}, extension of the fermionic and scalar sector of SM (IV)~\cite{Baker:2021nyl}. For such FOPTs we take benchmark values of $(\alpha, \beta/H, T)$ and  extract the PBH properties, i.e. the mass, abundance and initial rms spin. We list this benchmark points in Tab.~\ref{speccases} which shows that although the PBH spin are of the same order their magnitude is sensitive to the FOPT parameters.
We find that for physically realistic scenarios in various FOPTs, the spin is $\mathcal{O}(10^{-5})$. However, there exists various mechanisms, rooted in both astrophysics and particle physics, that can enhance the natal spin of PBHs. In the later, PBHs can spin up and have superradiant instability through emission of light axion or axion-like particles which are motivated from string axiverse~\cite{Calza:2023rjt,Calza:2021czr} whereas in the former, PBH mergers~\cite{Hofmann:2016yih} or close hyperbolic encounters~\cite{Jaraba:2021ces} can do the job. Though these processes are  in most cases model-dependent, but the final spin of the PBH almost inevitably depend on the initial spin. Depending on the specific mechanism, the final spin can be enhanced up to two orders of magnitude from the initial spin of PBHs created due to delayed vacuum decay during a FOPT. It is also relevant here that very light PBHs ($M_{\mathrm{PBH}}=\mathcal{O}(10^{15}\mathrm{~g})$), the abundance constraints are different depending on their spin~\cite{Dasgupta:2019cae}. Therefore this study, which gives a precise FOPT parameter dependence on spin, is imperative in order to get the evolution of PBH spin, from its production till today.
It is well established in literature~\cite{Saito:2023fpt} that phases of the universe, which were governed by equations of state softer than that of the radiation dominated universe, can be a breeding ground for highly spinning PBHs. Therefore, a study where FOPTs occur in the universe dominated by something other than radiation (e.g. early matter domination) can have huge phenomenological implications. On that note, it should also be taken into account that many of the considerations taken in this work, i.e., the $k$-dependence of the power spectrum, the threshold value of the overdensity, etc. would change in case of a universe governed by a different equation of state. This analysis  is out of the scope of this study and we leave it for future work.
\begin{table}[t]
\centering
\begin{tabular}{|c|c|c|c|c|c|c|}
\hline
BP & $\alpha$ & $\beta/H$ & \begin{tabular}[c]{@{}c@{}}$T$\\ (GeV)\end{tabular} & \begin{tabular}[c]{@{}c@{}}$M_{\mathrm{PBH}}$\\ (g)\end{tabular} & $f_{\mathrm{PBH}}$ & $10^{-5} \langle a_{*}^2\rangle^{1/2}$ \\ \hline \hline
I & 2.35 & 17 & $10^6$ & $9.23\times 10^{19}$ & 0.918 & 1.65 \\ \hline
II & 2.9 & 18 & 100 & $9.23\times10^{27}$ &0.0176  & 2.23 \\ \hline
III & 1.7 & 12 & 26 & $1.36\times10^{29}$ & 0.0161 & 2.04 \\ \hline
IV & 3.55 & 22 & $10^8$ & $9.23\times10^{15}$ & $0.00025$ & 1.17 \\ \hline
\end{tabular}
\caption{\footnotesize Benchmark points for BSM motivated FOPT parameters.}
\label{speccases}
\end{table}

%%%%%%%%%%%%%%%%%%%%%%%%%%%%%%%%%%%%%%%%%%%%%%%%%%%%%%
%%%%%%%%%%%%%%%%%%%%%%%%%%%%%%%%%%%%%%%%%%%%%%%%%%%%%%
\section{Summary and Conclusion}
\label{sec:concl}
%%%%%%%%%%%%%%%%%%%%%%%%%%%%%%%%%%%%%%%%%%%%%%%%%%%%%%
%%%%%%%%%%%%%%%%%%%%%%%%%%%%%%%%%%%%%%%%%%%%%%%%%%%%%%
In this article, we focus on peak theoretic analysis of the creation of PBHs during a FOPT due to the delay of vacuum decay and we obtain the delicate dependence of the spin and abundance of PBHs on the FOPT parameters, such as its strength $\alpha$, inverse of the characteristic time-span $\beta$, and its temperature $T$. We have also obtained the spin distribution of the PBHs for different combinations of these parameters. 

We find that $f_{\mathrm{PBH}}$ increases (decreases) steeply with increasing $\alpha$ ($\beta/H$), whereas, $\langle a_{*}^2\rangle^{1/2}$ increases (decreases) much slowly for the same. We  also show that the rms spin of a PBH population for realistic FOPT parameters is $\mathcal{O}(10^{-5})$, but the exact value of the spin is quite sensitive to the value of the parameters. One important point is that since the standard deviation of the overdensity is independent of the FOPT temperature, the spin also does not depend on the temperature. Next we have obtained the effect of the FOPT temperature of the PBH abundance for different combinations of $(\alpha, \beta/H)$. We found that for the same combination of $\alpha$ and $\beta/H$, the abundance increases with the FOPT temperature. In this article, we have ignored the parameter domain, where PBHs were being overproduced. For the distribution of PBH spin, we found that though the shape of the distribution remains qualitatively unaffected, quantitatively it varies with $(\alpha, \beta/H)$. We have considered a few benchmark FOPTs motivated by different BSM scenarios and show that the abundance and spin is very sensitive to the FOPT parameters.

It is worth mentioning that there are other prescriptions with similar PBH production mechanism of delayed vacuum decay where the standard deviation of the overdensity might be different functions of the FOPT parameters. Also, we have considered $\alpha\sim\mathcal{O}(1)$ and therefore, we have taken the critical, nucleation and percolation temperatures to be of the same order. But in cases of much larger $\alpha$ values, i.e. highly supercooled FOPT, this assumption would not hold true. Moreover, there are other model dependent mechanisms, e.g., Fermi balls, which could create PBH during a FOPT, but in these cases the implication of the peak theory might be much different as in these cases, the collapse happens due to the entrapment of fermions, as opposed to density perturbations which may lead to different initial spins. 
\\
\indent
Finally, we would like to emphasize that this analysis gives precise values of the PBH spin and abundance for a given combination of FOPT parameters. Even though the order of magnitude of this spin is small, there are mechanisms to enhance it to an appreciable level in which case there are multiple phenomenological implications.

\acknowledgments{IKB acknowledges the support by the MHRD, Government of India, under the Prime Minister's Research Fellows (PMRF) Scheme, 2022. We thank the anonymous referee for quite a few valuable suggestions.}

%\appendix

%\section{Appn2}
%\label{appn:2}
%%%%%%%%%%%%%%%%%%%%%%%%%%%%%%%%%%%%%%%%%%%%%%%%%%%%%%
%%%%%%%%%%%%%%%%%%%%%%%%%%%%%%%%%%%%%%%%%%%%%%%%%%%%%%
%%%%%%%%%%%%%%%%%   References   %%%%%%%%%%%%%%%%%%%%%
%%%%%%%%%%%%%%%%%%%%%%%%%%%%%%%%%%%%%%%%%%%%%%%%%%%%%%
\bibliographystyle{JHEP}
\bibliography{FOPT_PBH_SPIN_ref.bib}
%\include{bib}
%%%%%%%%%%%%%%%%%%%%%%%%%%%%%%%%%%%%%%%%%%%%%%%%%%%%%%
%%%%%%%%%%%%%%%%%%%%%%%%%%%%%%%%%%%%%%%%%%%%%%%%%%%%%%

\end{document}